\pdfoutput=1
\RequirePackage{tikz, pgfplots, pgfplotstable}
\usepgfplotslibrary{groupplots}

\documentclass[sn-standardnature]{sn-jnl}%

\jyear{2023}%

\theoremstyle{thmstyleone}%

\theoremstyle{thmstyletwo}%

\theoremstyle{thmstylethree}%

\begin{document}

\title[\texttt{APPFLx}: Secure End-to-End Federated Learning in Biomedical Research]{Enabling End-to-End Secure Federated Learning in Biomedical Research on Heterogeneous Computing Environments with \texttt{APPFLx}}

\author*[1]{\fnm{Trung-Hieu} \sur{Hoang}}\email{hthieu@illinois.edu}

\author[2]{\fnm{Jordan} \sur{Fuhrman}} %

\author[3]{\fnm{Ravi} \sur{Madduri}} 

\author[3,5]{\fnm{Miao} \sur{Li}} 

\author[3,4]{\fnm{Pranshu} \sur{Chaturvedi}} 

\author[3,4]{\fnm{Zilinghan} \sur{Li}} 

\author[3]{\fnm{Kibaek} \sur{Kim}} 

\author[6]{\fnm{Minseok} \sur{Ryu}}

\author[3]{\fnm{Ryan} \sur{Chard}} 

\author[3]{\fnm{E. A.} \sur{Huerta}}

\author[2]{\fnm{Maryellen} \sur{Giger}}

\affil*[1]{\orgdiv{Department of Electrical and Computer Engineering and Coordinated Science Laboratory}, \orgname{University of Illinois at Urbana-Champaign}, \orgaddress{\street{Street}, \city{Urbana}, \postcode{61801}, \state{IL}, \country{USA}}}

\affil[2]{\orgname{University of Chicago}, \orgaddress{\city{Chicago}, \state{IL}, \country{USA}}}

\affil[3]{\orgdiv{Data Science and Learning Division}, \orgname{Argonne National Laboratory}, \orgaddress{\city{Lemont}, \state{IL}, \country{USA}}}

\affil[4]{\orgdiv{Department of Computer Science}, \orgname{University of Illinois at Urbana-Champaign}, \orgaddress{\street{Street}, \city{Urbana}, \postcode{61801}, \state{IL}, \country{USA}}}

\affil[5]{\orgdiv{School of Industrial and Systems Engineering} \orgname{Georgia Institute of Technology}, \orgaddress{\city{Atlanta}, \state{GA}, \country{USA}}}

\affil[6]{\orgdiv{School of Computing and Augmented Intelligence} \orgname{Arizona State University}, \orgaddress{\city{Tempe}, \state{IL}, \country{USA}}}

\newcommand{\rv}[1]{\textsf{\textbf{\color{blue}{\tiny [Ravi: #1]}}}}
\newcommand{\hh}[1]{\textsf{\textbf{\color{magenta}{\tiny [Hieu: #1]}}}}
\newcommand{\mk}[1]{\textsf{\textbf{\color{cyan}{\tiny [Marcus: #1]}}}}
\newcommand{\newtext}[1]{\textcolor{blue}{#1}}
\definecolor{ForestGreen}{RGB}{34,139,34}

\abstract{Facilitating large-scale, cross-institutional collaboration in biomedical machine learning projects requires a trustworthy and resilient federated learning (FL) environment to ensure that sensitive information such as protected health information is kept confidential.
In this work, we introduce \verb|APPFLx|, a low-code FL framework that enables the easy setup, configuration, and running of FL experiments across organizational and administrative boundaries while providing secure \textit{end-to-end communication}, \textit{privacy-preserving} functionality, and \textit{identity management}. \verb|APPFLx| is completely agnostic to the underlying computational infrastructure of participating clients. We demonstrate the capability of \verb|APPFLx| as an easy-to-use framework for accelerating biomedical studies across institutions and healthcare systems while maintaining the protection of private medical data in two case studies: (1) predicting participant age from electrocardiogram (ECG) waveforms, and (2) detecting COVID-19 disease from chest radiographs. These experiments were performed securely \textit{across heterogeneous compute resources}, including a mixture of on-premise high-performance computing and cloud computing, and highlight the role of federated learning in improving model generalizability and performance when aggregating data from multiple healthcare systems. Finally, we demonstrate that \verb|APPFLx| serves as a convenient and easy-to-use framework for accelerating biomedical studies across institutions and healthcare system while maintaining the protection of private medical data.}


\maketitle
\setcounter{table}{1}

\section{Introduction}
\label{sec:intro}
In biomedical research, access to many types of data, such as identifiable electronic health records, medical images, and electrocardiogram (ECG) readings, is strictly regulated by law like HIPAA in the United States and GDPR in the European Union. Data access committees and institutional review boards (IRB) manually manage access controls to ensure that research is ethical and that the privacy of research subjects is safeguarded. Historically, these necessary, but cumbersome, access restrictions have had the undesirable effect of siloing data at the host institution which in turn has stymied collaborative research. Federated learning (FL)~\cite{pmlr-v54-mcmahan17a} has been proposed as one of the most viable frameworks to make protected health data available for training machine learning models across institutions without explicitly sharing any sensitive data  (privacy-preserving)~\cite{Kaissis2020SecurePA, Kaissis2021EndtoendPP, froelicher_truly_2021, bai_advancing_2021}. Such a distributed and privacy-preserving framework can address, and potentially transcend, the institutional restrictions limiting the exchange of data that is required to train cross-institutional machine learning models. 

Empowered by these capabilities, FL has recently garnered considerable attention in the research community and several solutions have been proposed~\cite{foley2022_openfl, beutel2020_flower, ryu2022_appfl, roth2022nvidia, web_ibm_fl}. For instance, Flower~\cite{beutel2020_flower} introduced a large-scale framework that can support up to 15 million parallel clients. While FATE~\cite{web_fate} focus on scalability and performance, suitable for industrial application, OpenFL~\cite{foley2022_openfl} positions itself as a flexible, extensible, and easily usable framework for data scientists.
Additional efforts have focused on ensuring data privacy using various privacy-preserving methods \cite{zhu2019_deep_leakage, geiping_inverting_2020-1, froelicher_truly_2021, wei2020_fedlearning}. In addition to these, we previously proposed \textit{Argonne Privacy Preserving Federated Learning} (\texttt{APPFL}) framework ~\cite{ryu2022_appfl}, a comprehensive end-to-end secure FL framework that includes multiple FL algorithms, differential privacy schemes, and communication protocols. Extensive experiments have been carried out to demonstrate its performance and communication efficiency on different biomedical datasets and computing environments~\cite{10.1088/1361-6560/acb754, yuexiang2023_fedscope}. 

\begin{figure}[t]
    \centering
    \subfigure[Workflow of cross-institute collaborative federated learning using \texttt{APPFLx.}]{
        \includegraphics[width=0.9\linewidth]{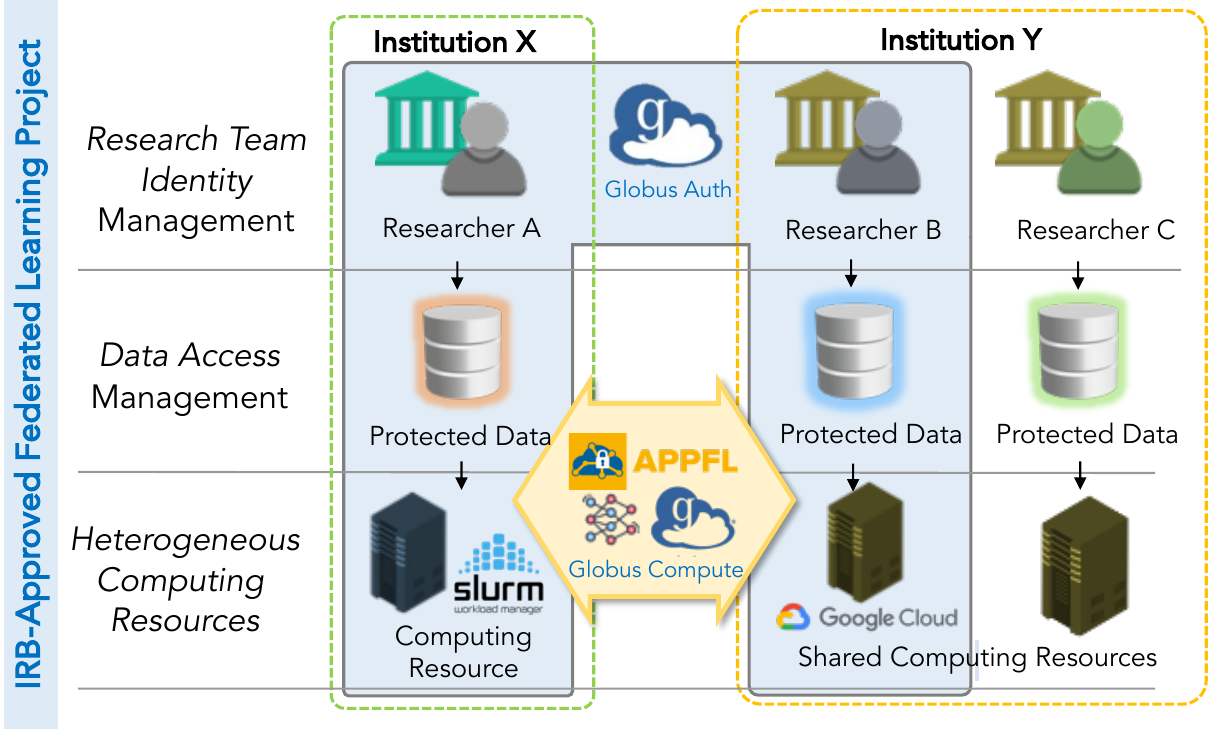}
        \label{fig:research_workflow}
    }

    \subfigure[\texttt{APPFLx} versus other federated learning frameworks.]{
        \includegraphics[width=0.7\linewidth]{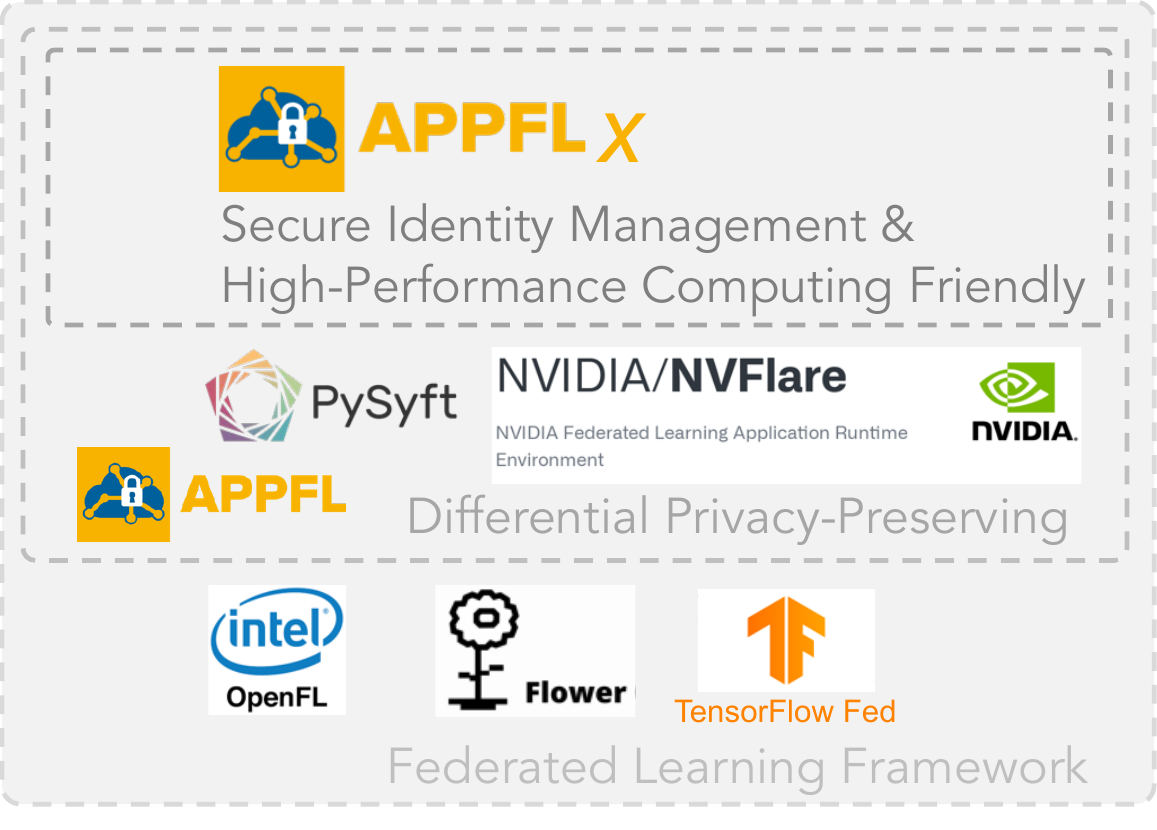}
        \label{fig:fed_framework}
    }

    \caption{(a) Our \texttt{APPFLx} enhances cross-institute collaborative federated learning workflow. Globus authentication service \cite{foster2011_globus, steven2016_globus_auth} and our \texttt{APPFLx} jointly provide researcher identity, data and computing resource management. In this diagram, while sharing the same computing and data storage resource at Institution $X$, only research $A$ and $B$ are approved to join the FL experiment. (b) Compared to other federated learning frameworks, \texttt{APPFLx} provides a comprehensive solution for biomedical research with differential privacy-preserving, secure identity management, and high-performance computing friendly.}
\end{figure}

\noindent \textbf{Challenges of FL Framework in Biomedical Research. } While much progress has been made, adopting a FL framework in cross-institute biomedical studies yields distinct challenges. First, under strict IRB regulations, it is essential to guarantee the trustworthiness of participating clients in the FL environment. Access should be restricted to authorized members only, and this requires a trusted identity and access management mechanism embedded into the framework. The second challenge is in handling heterogeneous compute --- such as on-premise supercomputing at one location, and cloud-computing services like Amazon Web Services or Google Cloud at another location --- that could also leverage different job schedulers, like Slurm \cite{yoo2003_slurm} or Load Sharing Facility (LSF) \cite{songnian1993_lsf} (\textbf{Figure~\ref{fig:research_workflow}}). A mature FL framework should address these challenges and be able to operate regardless of the underlying infrastructure and guarantee that access is limited to trusted partners. To address these challenges, we take advantage of \texttt{APPFLx}~\cite{li2023_appflx} that \textit{elaborates on the modular design and privacy-preserving aspects} of \texttt{APPFL}~\cite{ryu2022_appfl} with \textit{new capabilities} (\textbf{Figure~\ref{fig:fed_framework}}) which \textit{further eases the establishing of cross-institute biomedical research}. 

\noindent \textbf{Identity and Access Management in FL.} In order to ascertain that access is limited to trusted partners in the federation, there is a pressing need to authenticate users through their institutional identities that are integrated with their respective organizational Identity and Access Management (IAM) services. Globus Auth \cite{foster2011_globus, bryce2012_globus_ssas} provides one such secure service via a single sign-on solution, enabling researchers to easily access secure data and resources; this service is currently used in multiple large-scale initiatives across different scientific domains such as astronomy and genomics \cite{LAPLANTE2021100489, madduri2014experiences} . The Globus Auth service reduces the time and effort required to control access to data while ensuring the security and integrity of the shared resource. For these reasons, \textit{we integrated Globus Auth in \texttt{APPFLx} to set up secure federations using industry-standard identity and access management capabilities that enforce authentication and access control}. 

\noindent \textbf{FL on Heterogeneous High-performance Computing Resource.} 
In addition to the hard requirements of ensuring high-grade security, there are additional practical requirements needed to provide a simple-to-use framework.  For example, it is very likely that the participating institutions in the federation have heterogeneous computing capabilities such as on-premise high-performance computing or cloud computing facilities from different cloud providers. In an FL environment, there is generally a dedicated server that receives information from a series of participating clients that are collaboratively training a shared machine learning model. Clients independently perform local computations on their private data and communicate that result to the centralized server which in turn updates the joint model. We modeled these client-side training and updating steps as executing a federated function. As a result, we were able to leverage the Globus Compute~\cite{chard2029_funcx} service, that provides flexible distributed task execution mechanisms. This is enabled by exposing secure \textit{endpoints} on the participating client machines that are accessible to the centralized server. Globus Compute has been applied to high-performance computing tasks in many types of research \cite{huerta2021_accelerated, doi:10.1073/pnas.2100170118}. Since Globus Compute simply communicates data between a client and the server, i\textit{t makes the federation completely agnostic to the individual computing platforms of the participating clients}. This setup permits the client to leverage their institutionally available compute (e.g., number of compute nodes, cores per worker, allocation account) which removes the burden of managing system-detailed configurations on the centralized server resulting in considerably easier and faster deployment times. This is among the first implementations of Globus Compute as the communication backbone of an FL framework, together with FLoX~\cite{matt2023_flox}. Nonetheless, the focus of FLoX and \texttt{APPFLx} is different. FLoX is designed for single-user-multi-device cross-device FL settings, i.e., not relying too much on the Globus Identity Management, and all FL clients are different edge devices between the same person.

\noindent \textbf{Aims and Contributions.} 
Here, we describe the integration of Globus Auth~\cite{foster2011_globus} with Globus Compute~\cite{chard2029_funcx} into our previously proposed \texttt{APPFL} framework~\cite{ryu2022_appfl} to create \texttt{APPFLx} to expedite secure collaboration across multiple research institutions using protected health data. The main contributions of this study are provided as follows:

\begin{itemize}
    \item \texttt{APPFLx} streamlines the FL deployment process and provides both secure end-to-end communication and ensures strong identity and data access management controls that are compatible with organizational identity and access management services and with federal requirements. 
    \item A modular design that empowers users to develop customized FL algorithms that are agnostic to the underlying compute infrastructures.
    \item An FL-as-a-service web platform using \texttt{APPFLx} that enables non-experts to quickly set up and deploy secure federated learning experiments.
    \item To illustrate the pliability of this framework, we conducted extensive experiments on two real-world biomedical research domains in 1) estimating human biological aging from ECG waveforms and, 2) detecting COVID-19 disease from chest radiographs (CXR). The results demonstrate that federated learning studies can be easily established across four distinct research facilities to jointly train machine-learning models in a privacy-preserving fashion.    
    \item Finally, we also perform a model inversion attack to highlight the necessity of equipping differential privacy-preserving in FL and demonstrate the effect of applying the differential privacy protection scheme implemented by \texttt{APPFLx} framework. 
\end{itemize}

\section{Results}
\label{sec:case-studies}
In this section, we describe \texttt{APPFLx}  and results from secure, end-to-end, privacy-preserving FL in two distinct biomedical research studies. Particularly, we will describe the ease of setting up secure federations, designing and running FL experiments, and report the overall model performance. 

\subsection{Enabling secure federated learning in Biomedicine  using \texttt{APPFLx}}
\label{ssec:apptlx_biomedicine}
\noindent \textbf{Key Steps for Establishing FL Experiment. } \texttt{APPFLx} allows researchers to streamline an FL experiment in a simple 4-step process. We briefly summarize the workflow here:

\begin{enumerate}
    \item \textit{Identity Verification: } Participating institutions in FL sign-in to the federation using their institutional identity that is integrated with a Globus identity~\cite{foster2011_globus}. Globus Auth is integrated with identities of most US universities and national labs. The scientist that is setting up the FL (the \textit{orchestrator}) experiment organizes the participating Globus identities into a Globus group and assigns roles to each member, and thereby restricting the secure communication of data between only these parties. 
    \item \textit{System Setup: } All clients install and configure a Globus Compute-endpoint on the target compute and register the address of this endpoint with the centralized FL server.  
    \item \textit{Configuration: }The researcher can then define any model architecture and data loader in \texttt{PyTorch}~\cite{adam2019_pytorch} for their project. Additionally, training hyper-parameters, FL aggregation scheme, and privacy-preserving settings on the participating clients will be provided through a simple configuration file. 
    \item \textit{Running FL Experiment: } Finally, the researcher establishes and monitors the FL experiment. Besides regular training tasks, \texttt{APPFLx} also (optionally) performs cross-site validation, in which a model is sent and evaluated on the local private datasets at clients without sharing the data. 
\end{enumerate}

\noindent \textbf{Web-based User Interface for Rapid FL Setup. } In order to further lower the barrier in conducting FL experiments, we provide \texttt{APPFLx} as an FL-as-a-Service web platform \cite{li2023_appflx}. Using a user-friendly dashboard, a researcher can interactively complete all the required steps outlined above, such as managing Globus groups, registering new client endpoints, uploading model descriptions, and configuring FL parameters for both the server and the clients. The platform also has the ability to automatically initiate and deploy an \texttt{APPFLx} server, with appropriate access management policies expressed as AWS IAM rules, to an ECS container to act as the orchestration server. For started experiments, it is easy to monitor the training progress by leveraging existing visualization tools such as Tensorboard \cite{tensorflow2015-whitepaper} or inspecting the real-time client logs. The service is made publicly available at \url{https://appflx.link}.

In the following sections, we describe setting up FL experiments across organizational boundaries using heterogeneous computing capabilities for the training of ML models and fast iteration of FL experiments. We aim to demonstrate the flexibility and power of the \texttt{APPFLx} service while demonstrating the importance of FL in developing robust AI models under undesirable distribution shifts.

\subsection{Case Study 1: Biological Aging Prediction from ECG Signal}
\label{ssec:case_study_ecg}
\noindent \textbf{Task Description. } The electrocardiogram (ECG) is the most popular, simplest, and fastest exam used for the evaluation of various cardiovascular diseases. Predicting biological aging from the raw ECG waveform is beneficial for revealing an individual's cardiovascular health~\cite{lima_deep_2021, robyl2014_predicting}. In this study, we investigate the task of regressing human biological aging from 12-lead ECG waveform. A deep learning model is trained to reduce the mean squared error (MSE) between the predicted age and the subjects' aging at the date of ECG reading as ground-truth. 

\noindent \textbf{Datasets and FL Sites. }
This case study is a collaborative research project between Argonne National Laboratory \textit{(ANL)} and the Broad Institute \textit{(Broad)}. \textbf{Table}~\ref{tab:ecg-statistics} summarizes the statistics of the two datasets used at each client in the ECG experiment. At ECG-ANL, we adopted the publicly available PhysioNet dataset \cite{alday2020_physionet} while the dataset at ECG-Broad is composed of the ECG signals from the UK Biobank.

An FL experiment across multiple sites is established with \texttt{APPFLx}. In this study, a global server is hosted on a conventional CPU machine (specifically, Intel Core i7-6700K CPU @ 4.00GHz) at the University of Illinois at Urbana-Champaign (UIUC). The first client utilizes a computing cluster with a single NVIDIA GeForce RTX 3090, hosting and performing training on the ECG-ANL dataset. Meanwhile, the second client at Board Institute is initiated on a Google Cloud Compute Engine virtual machine instance which accelerates heavy-computing tasks with GPU power. This setup demonstrates the ability of \texttt{APPFLx} which favors a wide range of heterogeneous computing environments (e.g., local computing clusters and cloud computing services).

\begin{table}[t]
    \centering
    \subfigure[Statistics of the two datasets used in the ECG experiments.]{
        \label{tab:ecg-statistics}
        \begin{tabular}{c|ccc|c}
        \hline
        \textbf{Client} & \textbf{Train} & \textbf{Val} & \textbf{Test} & \textbf{Total} \\ \hline
        ANL          & 64518          & 7905         & 7905          & 80328 \\
        Broad        & 33140          & 4143         & 4143          & 41426 \\ \hline
        \end{tabular}
    }
    \subfigure[The mean square error of FL and local models.]{
        \label{tab:ecg-mse}
        \begin{tabular}{c|cc|c}
        \hline
        \multirow{2}{*}{\textbf{Training Dataset}}              & \multicolumn{3}{c}{\textbf{Testing Set}}                   \\ \cline{2-4} 
                                                                & \multicolumn{1}{c|}{\textbf{ECG-ANL}} & \textbf{ECG-Broad} & \textbf{Average} \\ \hline
        ECG-ANL \textit{(local training)}                               & \multicolumn{1}{c|}{109.95}   & 224.48  & 149.33   \\
        ECG-Broad \textit{(local training)}                             & \multicolumn{1}{c|}{225.41}   & 38.93   & 161.28   \\ \hline
        ECG-ANL+Broad - FedAvg\cite{pmlr-v54-mcmahan17a} & \multicolumn{1}{c|}{125.00}      & 41.70   & 96.35    \\ \hline
        \end{tabular}
    }
    
\caption{(a) Statistic of the datasets used in the biological aging prediction from ECG signal experiment. (b) Testing mean square error (MSE) of the biological aging prediction from ECG signal models. The average column computes the average MSE across two datasets, weighted by the number of testing samples.}
\end{table}

\noindent \textbf{Regression Model and FL Setup.} As a baseline model, we employed a ResNet34-styled~\cite{he2016residual} architecture. Each ECG channel is normalized to have zero mean and unit standard deviation. For the FL procedure, we use FedAvg~\cite{pmlr-v54-mcmahan17a} in our experiment for a total of $30$ federation rounds. At each training round, the local models of all clients are optimized in 2 local epochs before aggregating to the global model. We use Adam optimizer~\cite{kingma2015_adam} with an initial learning rate of 0.003 and a decay scheduler by a factor of $0.975$.

\noindent \textbf{Result - FL versus Local Training. } \textbf{Table} \ref{tab:ecg-mse} gives the MSE on the test set of models trained on individual (rows 1-2) and combined (row 3) datasets. We observe that \textit{models trained on the combined dataset remarkably outperform ones trained on the individual datasets} in average. This highlights the benefit of FL which allows training a machine learning model on the multiple datasets. Another noticeable outcome is that FL models typically achieve higher generalization rate when being cross-evaluated on unseen samples from out-of-distribution datasets.

\subsection{Case Study 2: COVID-19 Detection on Chest Radiographs}
\label{ssec:case_study_cxr}
\textbf{Task Description.} 
One of the most widespread implementations of Artificial Intelligence (AI) techniques for data that may require increased security or privacy is medical imaging.  However, allowing for open-source access to medical images can be cumbersome, particularly considering privacy concerns related to attack models and potentially inconsistent or ineffective data de-identification requirements across different local sites (e.g., face shearing technology).  The COVID-19 pandemic served as a prime example of a clinical use case for \texttt{APPFLx} implementation; many institutions were interested in contributing to aggregated datasets for use in developing medical imaging-based AI models for COVID-19 detection, differential disease diagnosis, and other radiological tasks. Despite good intentions, several complications arose through “Frankenstein” datasets, biased algorithms, and extensive time to open source \cite{driggs2021_pitfallsCovid}.  Many of these obstacles could have been alleviated if a privacy-preserving FL system were available. Thus, we consider COVID-19 detection on chest x-ray images (CXR) as a clinical use case to evaluate our proposed system.  Clinically, the most common methods of COVID-19 detection are non-imaging exams (e.g., antigen or RT-PCR exams), however, imaging could play a role in differential disease detection upon image acquisition in the future when non-imaging tests are less readily available or no longer standard practice.

\noindent \textbf{Datasets and FL Sites. } We set up a two-site training for this case study. 
The first site hosts the publicly available CXR dataset from the Medical Imaging and Data Resource Center \textit{(MIDRC)}.
Initiated in 2020 to combat the pandemic, MIDRC is a multi-institutional collaborative initiative in medical imaging through data sharing. This comprehensive dataset contains digital radiograph images, COVID test results, and demographic information collected from multiple hospitals.
The second one holds a private dataset that was collected at the University of Chicago  \textit{(UChicago)}.
The UChicago dataset is collected as part of the University of Chicago Center for Research Informatics (CRI) COVID-19 Datamart in conjunction with the Human Imaging ResearchOffice (HIRO). 
The two datasets' train-test splitting scheme, statistics, and the number of positive and negative samples used in this experiment are reported in \textbf{Table~\ref{tab:covid-statistics}}.

\noindent \textbf{Classification Model and FL Setup. } This case study demonstrates a simple transfer learning approach to CXR data. We fine-tune a ResNet18~\cite{he2016residual} model pre-trained on ImageNet~\cite{deng2009imagenet}. The last softmax layer is modified to match the binary classification task (i.e., COVID-19 positive and negative).
Further, other recent publications have investigated the relevance of \textit{personalized FL}~\cite{zying2022_personalized, viraj2020_survey}, or the development/improvement of FL models for performance at individual clients.  We incorporate personalized FL in this study through additional local \textit{fine tuning} on a small subset of labeled data (here, we conveniently adopt the validation set) at both the MIDRC and UChicago clients for an additional 40 epochs. Noteworthy, for this fine-tuning step, only the trainable parameters of the batch normalization layers~\cite{sergey2015_batchnorm, } are updated. This adjustment specifically targets the sensitivity of these layers to local data statistics. Similar strategies have been implemented in other domain adaptation approaches~\cite{li2017revisiting, wang2021tent}. 

Using \texttt{APPFLx}, we establish the federation across multiple institutes. For simplicity, we adopt the common FedAvg \cite{pmlr-v54-mcmahan17a} algorithm for the global aggregation. Two Globus Compute endpoints are installed at UChicago while the global server is hosted at UIUC. The clients securely store their dataset, and only reveal the data loaders to the server.  Once the federation is created, the server can automatically facilitate the training and cross-site evaluation process. During training, each client performs two local updates before being integrated into the global model. We repeat this process for a total of 40 global FL aggregation rounds. For hardware requirements, the center server only requires a CPU machine, while most of the heavy-computing tasks are performed locally on two GPU clusters of the University of Chicago (HPE Superdome Flex NUMA computation server with 2 NVIDIA Tesla V100 32GB GPUs).  

\noindent \textbf{Result - FL versus local training.} All models are evaluated on the clients' test set using the cross-site validation feature of \verb|APPFLx|. We separately plot the ROC curves (\textbf{Figure} \ref{fig:covid-roc}-left), confusion matrices (\textbf{Figure} \ref{fig:covid-roc}-right), and compute the corresponding areas under the ROC curve (AUC), confidence interval (\textbf{Table} \ref{tab:covid-auc-score}) when testing them on the two datasets. 
We compute the AUC as the main evaluation metric for discussion. 
Over the two datasets, we observe that while models trained on a single dataset (rows 1-2) achieve satisfactory performance when evaluating on test sets with the same distribution, the performance\textit{ drops significantly on the test set of other sites} (e.g., $0.80 \rightarrow 0.56$ for the model trained on MIDRC dataset and $0.67 \rightarrow 0.56$ for the model trained on UChicago). Meanwhile, the model trained in FL settings with combined MIDRC+UChicago dataset (row 3) can overcome this circumstance on single-dataset models, and provide a more stable performance. Notably, for the MIDRC dataset, the performance of the FL model ($0.69$) was significantly reduced in comparison to the local model ($0.80$). This result indicates joint training with the UChicago dataset with a large distribution gap can possibly downgrade the performance of FL model. 

However, our investigation of \textit{fine tuning }for personalized FL through \texttt{APPFLx} (row 4) demonstrated \textit{superior performance}, achieving comparable performance to the locally-trained MIDRC model and significantly outperforming the local UChicago model. This suggests that the baseline model provided by the FL algorithm serves as a better foundation than the original ImageNet-trained~\cite{deng2009imagenet, he2016residual} model for addressing our task of COVID-19 detection.  This has potential practical implications in how FL is deployed, particularly in scenarios with differing data distributions across FL clients as in our case study; specifically, in such scenarios when a single model cannot effectively achieve high performance simultaneously across clients, personalized FL as deployed in our \texttt{APPFLx} scheme could play a significant role in the development of improved models.

\begin{table}[t]
  \subfigure[Statistics of the two datasets in the COVID-19 detection on CXR case study.]{
    \label{tab:covid-statistics}
    \centering
    \resizebox{\linewidth}{!}{
    \begin{tabular}{c|ccc|c}
    \hline
    \textbf{Client} & \textbf{Train} & \textbf{Val} & \textbf{Test} & \textbf{Total} \\ \hline
    MIDRC           & 9867 (4226+/5641-)      & 2056 (925+/1131-)     & 2081 (932+/1149-)           & 14004\\
    UChicago        & 26047 (4226+/23226-)          & 5569 (587+/4982-)      & 5619 (637+/4982-)         & 37235 \\
     \hline
    \end{tabular}
    }
  }

\subfigure[AUC comparison between local and FL models.]{
    \label{tab:covid-auc-score}
    \centering
    \resizebox{\linewidth}{!}{
    \begin{tabular}{c|c|c}
        \hline
        \multirow{2}{*}{\textbf{Training Dataset}}                   & \multicolumn{2}{c}{\textbf{Testing Set}}  \\ \cline{2-3} 
                                                                     & \textbf{MIDRC} & \textbf{UChicago} %
                                                                     \\ \hline
        MIDRC \textit{(local training)}                              & 0.80 [0.78, 0.82]    & 0.56 [0.54, 0.58]    %
        \\
        UChicago \textit{(local training)}                           & 0.59 [0.57, 0.62]    & 0.67 [0.65, 0.69]    %
        \\ \hline
        \multicolumn{1}{l|}{MIDRC+UChicago - FedAvg \cite{pmlr-v54-mcmahan17a}}   & 0.69 [0.67, 0.71]    & 0.67 [0.65, 0.69]  %
        \\
        MIDRC+UChicago - FedAvg + Fine Tuning                        & 0.79 [0.77, 0.80]    & 0.84 [0.83, 0.84]    %
        \\ \hline
        
    \end{tabular}
    }
}
\caption{(a) Statistics of the datasets used in the COVID-19 chest X-ray image recognition experiment. Numbers in the parentheses indicate the number of positive $(+)$ and negative $(-)$ samples. (b) AUC score of the COVID-19 chest X-ray image recognition models. Statistical significant comparisons within a testing population are bold.} 
\end{table}

\input{figures/cfs_roc_plots}

\subsection{Model Inversion Attacks and Privacy-Preserving Demonstration on Case Study 2}

Privacy-preserving, an essential part for a trustworthy and secure FL, is one of the most important component in our \texttt{APPFLx}. In this section, we take the preceding Case Study 2 (\textbf{Section.}~\ref{ssec:case_study_cxr}) to demonstrate the risk of user privacy leakage via an inversion attack simmulation, and the advantage of \texttt{APPFLx} privacy-preserving scheme to mitigate this critical issue.

\noindent \textbf{Model Inversion Attacks and Privacy-Preservation.} 
Recent works~\citep{zhu2019_deep_leakage,zhao2020idlg, geiping_inverting_2020-1,huang2021evaluating,yin2021see,geng2023improved} have examined a scenario where an attacker has access to the \textit{local client gradient} updates during FL processes. Alarmingly, the gradients information communicated between the clients and the server during training or even after training can reveal information of the training set in which an iterative gradient inversion algorithm can reasonably reconstruct private local data. 
Thus, we evaluate the privacy-preserving scheme implemented by the \texttt{APPFLx} framework to qualitatively and quantitatively measure  the extent of information leakage through gradients by performing an inversion attack during the FL process with and without our privacy-preserving scheme. 

\noindent \textbf{Experimental Setup. }
Following the same setting, we study a ResNet18~\cite{he2016residual} model pre-trained on ImageNet~\cite{deng2009imagenet}  which consumes a grayscale input image of size $224\times224$ as input. For a fair comparison with previous work~\cite{Kaissis2021EndtoendPP}, the publicly available dataset~\citep{OMS} is used in our inversion attack experiments. Other setups follow the setting described in \cite{zhu2019_deep_leakage}, and the effect of differential privacy when varying the clipvalue ($c$), noise level ($\epsilon$), amount of training, and the training batch size ($b$) on the reconstructed image is investigated.

\input{figures/inversion_attack}
\noindent \textbf{Result - Model Inversion Attack. } 
\label{ssec:result_inversion}
Our model inversion attack results with \textit{increasing differential privacy preserving budget} are illustrated in \textbf{Figure}~\ref{fig:reconstruction_qualitative}-top.
(a) provides the baseline result (described in \textbf{Section}~\ref{ssec:method_inversion_attack}), as the most vulnerable case, and without our privacy-preserving scheme on a single-sample training dataset with batch size $b = 1$. (b) uses the same setting as (a) except Laplacian noise with clip value $c = 1$ and $\epsilon = 0.1$ is added to tackle the inversion attack (privacy-preserving scheme). 
We observe that reconstruction quality is significantly downgraded compared to (a). 
In (c) and (d) we further increase the Laplacian noise (by decreasing $\epsilon = 0.05$ to $\epsilon = 0.01$), respectively. Here, the reconstructed image is completely unrecognizable as CXR.
For quantitative evaluation, the mean squared error (MSE) and peak signal-to-noise ratio (PSNR) for the increasing privacy-preserving levels (top row, right bar charts) showcases a \textit{clear decline in reconstruction quality}, indicating the benefits of \textit{incorporating the privacy-preserving scheme} of \texttt{APPFLx}.

Meanwhile, (\textbf{Figure}~\ref{fig:reconstruction_qualitative}-bottom) studies the effects of other training factors on the inversion attack \textit{without} applying the differential privacy scheme. In (e) and (f), the light training and extra training of the same ResNet 18~\cite{he2016residual} model are compared. 
In the light training mode, the model explores 20 images while this number increases to 150 in the extra training case. In both scenarios, the amount of gradient information leakage is still remarkable (in contrast with (c) and (d), (e) and (f) have comparatively low MSE and high PSNR values, and only worse than (a)). 
Additionally, we further study the effect of increasing training batch size with $b = 10$ (g) and $b = 50$ (h).
The MSE and PSNR metrics (bottom row, right bar charts) verify the effect of increasing batch size. 
Surprisingly, in both cases, the image reconstruction is still recognizable given that the gradient update from a large batch of images contains much less image-specific information than a single-image batch (as in (a))
To sum up, these results indicate that increasing the amount of training or changing the batch size, up to some extent, can reduce the reconstruction quality. However, they \textit{can not completely eliminate the role of privacy-preserving scheme} and privacy-preserving is \textit{always important at every stage} of the training process.

\section{Discussion}
\noindent \textbf{Rapid FL Experiment Deployment.} 
Far as we are aware, \texttt{APPFLx} is \textit{secure} and \textit{compatible} with the workflow of research organizations (\textbf{Section}~\ref{ssec:apptlx_biomedicine}). In the two case studies (\textbf{Section}~\ref{ssec:case_study_ecg},~\ref{ssec:case_study_cxr}), we showcase a rapid incorporation for conducting FL experiments across four research institutes (ANL, UChicago, UIUC, and Broad). There is no trade-off when using \texttt{APPFLx} compared to other popular FL frameworks. In terms of functionality, there are no additional technical set-up steps required. The framework provides similar training utilities, and privacy-preserving ability, and can be instantaneously deployed.

\noindent \textbf{Noticing the Cross-dataset Generalizability with FL.} Understanding model generalizability is an important task in machine learning. Using \texttt{APPFLx}, we demonstrated this process in action by assessing the single-dataset model's generalizability on \textit{cross-site evaluation} and how FL comes into place in two real-world case studies. While the benefit of FL in improving generalizability has been shown in previous research, our result further highlights the \textit{need to adopt FL techniques} when experimenting with machine learning models in biomedical research. \texttt{APPFLx}, flexible design framework, becomes a convenient tool for researchers to quickly set up FL experiments, analyzing and improving the model generalizability across multiple sites.

\noindent \textbf{FL on Clients with Severe Distribution-shift.} Utilizing a large pool of training data, FL models are \textit{typically expected} to be well generalized and capable of achieving supreme performance on datasets from both cross-site and same-site distributions. However, with large distribution-shift, and a limited number of participating FL sites demonstrated in our case study, their performance still lags behind the locally trained models when evaluating on same-site testing sets. Notice that this is a common phenomenon in FL, which is even more likely to happen in biomedical research. Fortunately, this shortcoming has been partially addressed in several previous personalized FL research \cite{yiquiang2020_fedhealth, jiang2023testtime, viraj2020_survey, zying2022_personalized}, and in the Case study 2 (\textbf{Section}~\ref{ssec:case_study_cxr}), a fine-tuning scheme has been adopted to overcome this drawback.

\noindent \textbf{Comparison to Prior Work}. From the aspect of identity management, many FL frameworks~\cite{foley2022_openfl, ryu2022_appfl} verify digital entities involved in FL process using a common trusted certificate authority (CA) on an HTTPS server for listening and signing requests. Other methods also extend it into a blockchain-based and decentralized system using smart-contract~\cite{jiahui2021_did-efed}. While these approaches can verify the ownership of a public key by a named subject, there is a lack of correspondence between virtual and real-world identity. Our \texttt{APPFLx} takes a step further by coupling digital FL clients with researchers' identities. Note that this single-time authentication governs all aspects of an FL process, including granting access to sensitive data, and allocating computing resources, which are naturally controlled by the researcher’s identity. To the best of our knowledge, this is the first FL framework that enables this ability.

Most of the aforementioned FL frameworks typically target edge devices~\cite{alex2022_fl_edge} as their main deployment target. Some frameworks \cite{web_ibm_fl, web_nvidia_clara, roth2022nvidia} support leveraging GPUs for hardware acceleration which supports the development of FL applications that require real-time response and cater to applications that compute intensive. However, there is an ongoing need to develop FL technologies that can leverage heterogeneous high-performance computing (HPC) resources, similar to \cite{web_ibm_fl, web_nvidia_clara}. We address these requirements with Globus Compute integration to the \texttt{APPFL} toolkit. The Globus Compute integration allows to \textit{seamless} leverage of traditional tightly-coupled high-performance computing resources and cloud computing resources that increase the overall adoption scenarios.

\noindent \textbf{Privacy-preserving for FL training. }  Qualitative and quantitative results across all three inversion attack experiments show that adopting a differential privacy scheme is crucial all the time. Using a simple attack model on the leakage of gradient communication between server and client can easily violate the privacy of user data. We see that in a realistic setting, where the data batch size gets larger and after some training has been conducted on the target network, stealing private training information from clients is still possible. Fortunately, with a differential privacy scheme, even the most susceptible setting of batch size one and using one training image only can tear down the inverse relation of gradient sent by the client and the underlying training data, thus eliminating the chance of data leakage. The capability of \texttt{APPFLx} framework allows all FL participants to enable the privacy-preserving scheme when needed, creating a\textit{ trustworthy FL environment for conducting cross-institute biomedical research}.

\noindent \textbf{Limitations and Future Work. } The initial setup of training sites for clients requires experience in configuring Globus Compute endpoints and going through several platforms like Globus, which may be an obstacle in some situations. Client-server communication is tightly intertwined with Globus Compute, creating a substantial dependence on this service and may limit the scalability of \texttt{APPFLx} when increasing the number of clients. While Globus has been adopted by several organizations, some organizations may be restricted to using products that use more common authentication standards. Future work includes exploring alternative authentication standards prevalent in the industry such as OAuth2 or OpenID. In addition, a more complete investigation of the system's security is necessary to recognize possible system vulnerabilities and thoroughly assess the privacy-preserving capabilities of \texttt{APPFLx}. 

\noindent \textbf{Conclusion. }
In this paper, we propose the integration of Globus compute to \texttt{APPFL}, namely \texttt{APPFLx}. This is a free, open-source high-performance federated learning platform that is specialized for facilitating cross-institutional collaboration on sensitive medical data. The two key novelties of our framework include \textit{organizational identity management} and \textit{task execution strategy on heterogeneous computing resources}. We further demonstrate the use of \texttt{APPFLx} in two real-world case studies: biological aging prediction from  ECG signal and COVID-19 chest x-ray image recognition. Our framework successfully coordinates a federated learning process across four research institutes.

\section{Methods}
\subsection{Architecture of \texttt{APPFLx}}
\label{sec:appfl-funcx}
\begin{figure}[t]
    \centering
    \includegraphics[width=\linewidth]{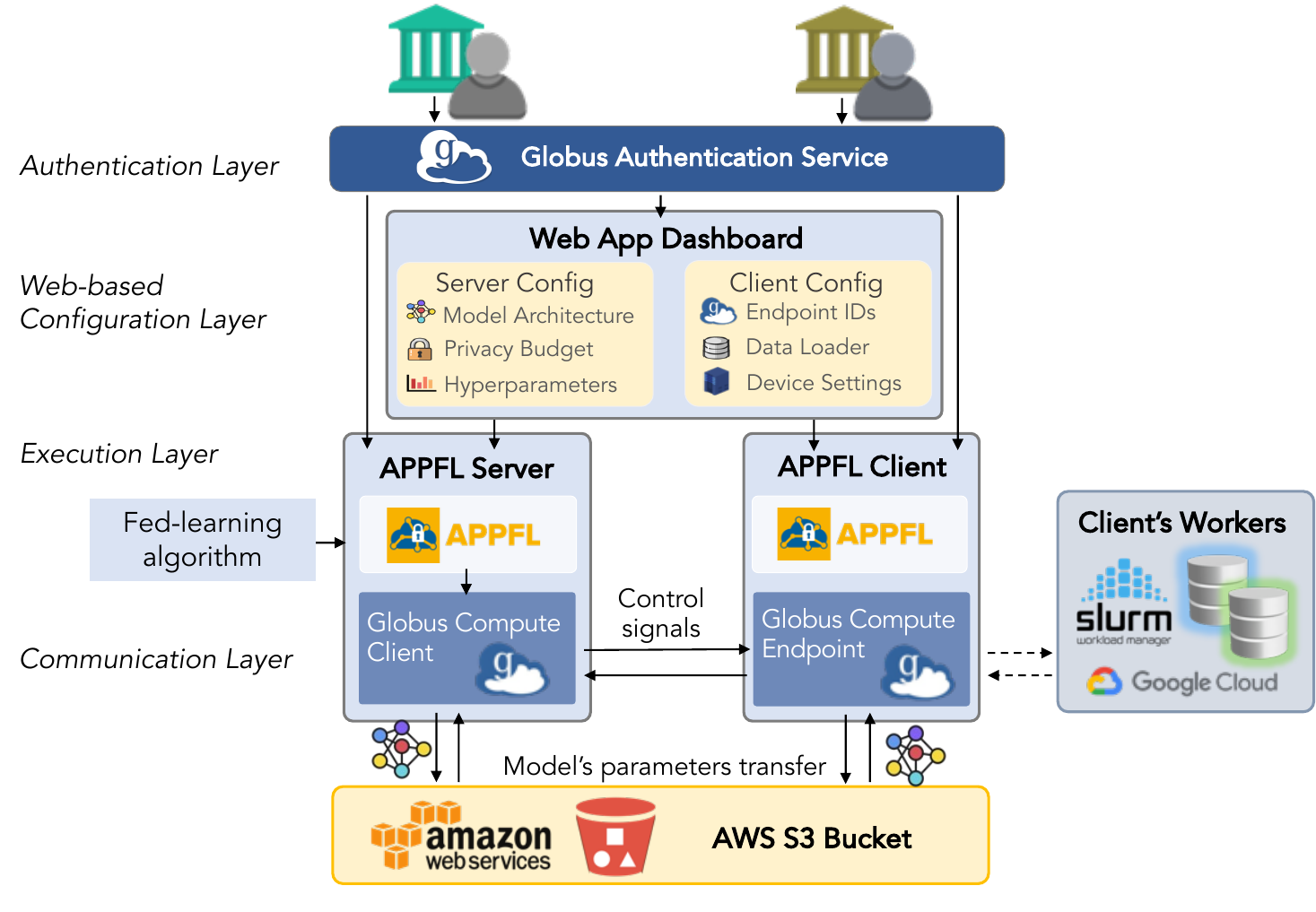}
    \caption{Main components of the \texttt{APPFLx}. Modules can be classified into four layers: authentication, web-based configuration, execution and communication.}
    \label{fig:main_components}
\end{figure}

Starting with \texttt{APPFL}~\cite{ryu2022_appfl}, we then integrate Globus Compute~\cite{chard2029_funcx} to establish a mechanism for handling remote task execution, identity management, and performing federated learning at multiple distributed clients. Figure \ref{fig:main_components} illustrates the main components of our system, classified into four main layers, namely \textit{authentication, web-based configuration, execution} and \textit{communication} layer. 

\noindent \textbf{Authentication Layer. }
Globus Authentication (Globus Auth) \cite{steven2016_globus_auth} - a trustworthy service for end-user identity management is used extensively in our framework. GA serves as a broker for authentication, identity provider, and managing interaction between end-users. After the authentication step, all verified credentials will be passed through the execution layer where the workers execute the training process on behalf of the researcher. An illustration is given in \textbf{Figure}~\ref{fig:research_workflow}, where the researcher \textit{A} and \textit{B} are grouped together when involved in the same project using GA while \textit{C} does not. In this experiment, \textit{A} and \textit{B} have the right to access their institution's computing and data management resources. During the FL, \texttt{APPFLx} acts as a proxy for the two researchers, where it is now authenticated to have the same right as its owner to perform FL tasks.

\noindent \textbf{Web-based Configuration Layer. }
The purpose of the web application is to streamline the configuration process for users and enhance the user experience. We used Python Flask framework~\cite{grinberg2018flask} for setting up the backend, and Globus~\cite{foster2011_globus} for authentication. Using a unified dashboard, users can provide server-side and client-side configurations. Client-side configuration includes data loaders, customized for their training data, and an identity string of Globus Compute-endpoint for registering their computing resource. The group administrator can then initiate the server-side settings, which include a list of registered clients, the model architecture, FL algorithm, and specify the privacy-preserving budget and other training hyperparameters. We also provide some helper functions that help users check the clients' status and aggregate training logs from either a regular text file or Tensorboard. Optionally, users can select an automated \texttt{APPFLx} server deployment on an ECS container(provided by AWS \cite{web_aws}), to quickly set up an experiment. 

\noindent \textbf{Execution Layer. }
The main challenge when working with multiple HPC systems is the heterogeneity of their cluster management and job scheduler. HPC administrators may or may not adopt a platform like Slurm~\cite{yoo2003_slurm}. To successfully execute the jobs, users also need to specify worker configurations (e.g., default job queue, charging budget, etc.) which creates a great burden on the FL framework. To this end, we utilize the advantage of Globus Compute~\cite{chard2029_funcx} and make it becomes the main backbone for network communication and executable jobs management. The benefit of using Globus Compute is that this burden of handling dispatch job execution over diverse platforms is done automatically through a universal programming interface. Globus Compute also favors various a wide range of computing resources, from personal computers to supercomputing facilities, like the Delta supercomputer (NCSA), or Theta (ALCF). Another benefit of Globus Compute is that the HPC configuration is dedicated to the client's owner when setting up the \texttt{APPFLx} clients which removes the need to continuously communicate with the server when updating configurations.

We briefly summarize how Globus Compute is adopted in \texttt{APPFLx}. Globus Compute consists of two main components: Globus Compute-Client and Globus Compute-endpoint, which are deployed at \texttt{APPFLx} server and client, respectively. Endpoints are abstract representations of computing resources that run in the background on the client's machine on behalf of the authenticated user and handle dispatch job execution from funcX clients on request. Meanwhile, Globus Compute client at \texttt{APPFLx} \textit{server} serves as a controlling node for managing a set of distributed \texttt{APPFLx} clients, sequentially assigning training tasks during an FL experiment. We require all servers and clients must have an outbound Internet connection.

\noindent \textbf{Communication Layer. }
While Globus-computing can effectively communicate the structure of machine learning models, and training configurations from server to all clients, it is not optimized for transferring large binary files, such as model weights. To this end, we adopt the S3 Bucket provided by Amazon Web Service (AWS) \cite{web_aws} to accommodate this task.

\subsection{Model Inversion Attack Experiment}
\label{ssec:method_inversion_attack}
\noindent \textbf{Inversion Attack via Gradient Data Leakage.} According to \cite{geiping_inverting_2020-1,Kaissis2021EndtoendPP,hatamizadeh2023gradient}, gradient data from the early iterations of training are generally \textit{more susceptible to training data leakage}. This can be explained by the fact that during training, the magnitude of the gradient updates generally converges to zero regardless of the underlying training data. 
Because of this tendency, at later training steps, the gradient may only contain a small amount information of from the training sample while the earlier ones contain a sufficient amount of information to reconstruct the private training data on the client (since the gradients are generally still far from zero). 
Therefore, with a fixed number of training epochs, we investigate different dataset sizes as larger datasets require more training steps of the neural networks. 

Meanwhile, larger training batch size can also adversely impact the reconstruction quality as the gradient update representing a batch of training examples is an average of the training batch. 
Hence, inspecting an individual gradient corresponding to each training image for a single-image inversion attack in a large training batch is more challenging. 
Previous works have shown that both the amount of training and batch size reduces the efficacy of the gradient inversion attack \citep{zhu2019_deep_leakage,zhao2020idlg,geiping_inverting_2020-1,huang2021evaluating,yin2021see,geng2023improved}.

\noindent \textbf{Inversion Attack Baseline}. The most susceptible scenario for an inversion attack is when the training dataset size, training batch size, and number of training rounds are all set to one (referred as \textit{baseline} attack). Hence, this setup is utilized to compare the effects of increasing the level of differential privacy, amount of training, and batch size (the results are provided in \textbf{Section.}~\ref{ssec:result_inversion}). 
When varying the level of differential privacy, the clip value of the Laplacian mechanism~\cite{dwork2006calibrating} is set to one while the value of $\epsilon$ \textit{gradually decreases} (increasing the scale of the Laplacian noise, equivalently).
When examining the influence of the training amount, we train the network model using 20 images and then perform a gradient inversion attack over this \textit{lightly trained} model. The same process is repeated with 150 images in the \textit{extra training} case. 
Finally, regarding the effect of training batch size, we compare the reconstruction results for \textit{various batch size choices} (1, 10, and 50).

\noindent \textbf{Details of the Attack Implementation.} 
We use the inversion attack model given by \cite{geiping_inverting_2020-1} with an additional modification which includes a batch normalization penalty introduced in \cite{yin2021see}. 
The general strategy is to initialize the inversion attack algorithm with a placeholder image, which will be updated continuously through an iterative inverting gradient algorithm and eventually converged to the private training data sample.  
Four different choices of initialization are experimented with: sampling pixel values from a random Gaussian distribution~\cite{geiping_inverting_2020-1}, uniform distribution~\cite{Kaissis2021EndtoendPP}, and creating a new image by averaging several images over a non-overlapping data set~\cite{hatamizadeh2023gradient}. 
We also investigate two common optimizer choices for the attack algorithm, including Adam\cite{DBLP:journals/corr/KingmaB14} (used in \cite{geiping_inverting_2020-1}) and AdamW\cite{DBLP:conf/iclr/LoshchilovH19} (used in \cite{Kaissis2021EndtoendPP}).
Extensive grid search amongst all combinations of the aforementioned image initialization, scale of total-variation penalty~\cite{geiping_inverting_2020-1}, BN penalty \cite{yin2021see} is necessary for each setup to produce pronounced reconstruction results.

\bibliography{sn-bibliography}%

\end{document}